\documentclass[11pt,a4paper]{article}
\usepackage{amsmath,amssymb,amsfonts,slashed}
\usepackage[text={176mm,240mm},centering]{geometry}
\usepackage{float}
\usepackage[figuresright]{rotating}
\usepackage{multirow}
\usepackage{color}
\usepackage[colorlinks=true,linkcolor=blue,breaklinks=true,urlcolor=blue,citecolor=blue]{hyperref}
\usepackage{cite}

\newcommand \bra [1] {\langle {#1}\vert}
\newcommand \ket [1] {\vert {#1}\rangle}

\def\be{\begin{equation}}
\def\ee{\end{equation}}
\def\bal#1\eal{\begin{align}#1\end{align}}
\def\PPP/{${}^3P_0$}
\def\cce/{coupled-channel effects}
\def\doned/{$D_1\bar{D}$}
\def\dmeson/{charmed meson}
\def\swave/{$S$ wave}
\def\dwave/{$D$ wave}
\def\xy/{$X(4260)$}
\begin{document}
\title{$X(4260)$ Revisited: A Coupled Channel Perspective}

\date{\today}

\author{
     Yu Lu$^{1,3}$
     Muhammad Naeem Anwar $^{2,3}$
     Bing-Song Zou$^{2,3}$
       \\[2mm]
      {\it\small$^1$Institute of High Energy Physics, Chinese Academy of Sciences, Beijing 100049, China,}\\
      {\it\small$^2$CAS Key Laboratory of Theoretical Physics, Institute of Theoretical Physics,}\\
      {\it\small  Chinese Academy of Sciences, Beijing 100190, China}\\
      {\it\small$^3$University of Chinese Academy of Sciences, Beijing 100049, China}\\
}

\medskip
\maketitle

\begin{abstract}
We calculate the probabilities of various charmed meson molecules for \xy/ under the framework of the \PPP/ model.
The results indicate that, even though heavy quark spin symmetry forbids
$S$ wave coupling of \doned/ to the ${}^3S_1$ charmonia [$\psi(nS)$],
the $D$ wave coupling is allowed and not negligible.
Under this symmetry, the \doned/ can couple to ${}^3D_1$ charmonia [$\psi(nD)$] via both $S$ and $D$ waves,
and the overall coupling is around three times larger than that of $\psi(nS)$.
The \xy/ cannot be a pure molecule but a mixture of a charmonium and various charmed meson components.
Since the \doned/ couples strongly to $\psi(nD)$,
our results suggest that, in the \doned/ molecular picture,
the charmonium core of \xy/ is $\psi(nD)$ instead of $\psi(nS)$.
As a result, the experimental fact that the $R$ ratio has a dip around 4.26~GeV can be understood in the \doned/ molecular picture of the \xy/.
\end{abstract}

\section{Introduction}

In 2005, the \emph{BABAR} collaboration~\cite{Aubert:2005rm} found a peak around
$4259~\text{MeV}$ in the initial state radiation (ISR) process $e^+e^-\to \gamma_{\text{ISR}} J/\psi \pi \pi$.
Since this resonance is directly generated via $e^+e^-$ annihilation, the $J^{PC}$ should be $1^{--}$.
This resonance is later confirmed by the CLEO and Belle Collaborations\cite{Coan:2006rv,Wang:2007ea,Yuan:2007sj}
and now is known as \xy/ [previously named as $Y(4260)$].
Since the \xy/ is far above the $D\bar{D}$ threshold,
it has a large phase space to decay into \dmeson/ pairs.
Nevertheless,
no open-charm decay has been observed up to now\cite{Abe:2006fj,Pakhlova:2008zza,Pakhlova:2007fq,Pakhlova:2009jv,Aubert:2006mi,Aubert:2009aq,CroninHennessy:2008yi}.
The mass and the unique decay patterns of \xy/ stimulate numerous theoretical studies.

Some people try to accommodate \xy/ in the potential quark model\cite{LlanesEstrada:2005hz,Li:2009zu}.
For example, Llanes-Estrada claims that it is dominantly $\psi(4S)$  in the relativistic quark model\cite{LlanesEstrada:2005hz},
and in order to explain the small cross section in the $e^+e^-$ collider,
an $S-D$ mixing mechanism is also introduced.
Apart from this charmonium scenario,
most people suggest it to be a noncharmonium candidate,
including four-quark state\cite{Maiani:2005pe,Ebert:2005nc,Ebert:2008kb},
$c\bar{c}$ hybrid\cite{Zhu:2005hp,Kou:2005gt,Close:2005iz,Chen:2016ejo},
different molecular scenarios,
such as $D_1(2420)\bar{D}+c.c.$ (or \doned/ for short) \cite{Li:2013bca,Wang:2013cya,Wang:2013kra,Qin:2016spb},
$D_0\bar{D}^*+c.c.$\cite{Ding:2008gr}, $\rho^0 \chi_{c0}$\cite{Liu:2005ay}, $\omega \chi_{c1}$\cite{Yuan:2005dr},
$\omega \chi_{c0}$\cite{Dai:2012pb}, and $J/\psi K\bar{K}$\cite{MartinezTorres:2009xb},
or even nonresonance interpretation\cite{Chen:2010nv}.
For more information,
we suggest following the review papers\cite{Brambilla:2010cs,Chen:2016qju,Guo:2017jvc}.

One of the reasoning in the \doned/ molecular scenario is that the \xy/ is very close to the \doned/ threshold.
However, this scenario is challenged by Li and Voloshin who claim that,
with respect to the heavy quark spin symmetry (HQSS),
\doned/ cannot couple to ${}^3S_1 \ c\bar{c}$ via an \swave/,
and thus its production in $e^+e^-$ collisions should be heavily suppressed\cite{Li:2013yka}.
However, Wang \emph{et. al}~\cite{Wang:2013kra} discuss the HQSS breaking effects due to the finite mass of charm quark which make the \doned/ molecule explanation reasonable for \xy/,
and the claimed suppression is even welcome to explain the nonobservation of the \xy/ in the $R$-ratio scan.

In both papers\cite{Li:2013yka,Wang:2013kra}, the $D$ wave \doned/ contribution is neglected at the hadronic level.
In the quark model,
the wave functions encapsulate both long and short distance information,
and one can use them to revisit the above conclusions at the quark level.
However, so far, there is no such explicit calculation.
We do not aim to fully explain the mass and decay properties of the \xy/ in this paper,
instead, we will study the \dmeson/ components in the \xy/,
and, in particular, analyze the prerequisites of the \doned/ molecule scenario.

This paper is organized as follows.
In section \ref{sec:ppp}, we give a brief introduction of the calculation framework,
including the potential model, the \cce/, \PPP/ model and some subtleties from the HQSS, and the wave functions.
Section \ref{sec:result} is devoted to the analysis of the results.
Finally, we give a short summary in Sec.~\ref{sec:summary}.

\section{Calculation Framework}\label{sec:ppp}

In the quenched quark model,
the wave functions for charmonia are obtained by solving the Schr\"{o}dinger equation
with the well-known Cornell potential~\cite{Eichten:1978tg,Eichten:1979ms}
\be
V(r)=-\frac{4}{3} \frac{\alpha}{r}+\lambda r+c,
\ee
where $\alpha, \lambda$ and $c$ stand for the strength of color Coulomb potential,
strength of linear confinement and mass renormalization, respectively.
The hyperfine and fine structures are generated by the spin-dependent interactions
\be
V_{s}(r)=\left(\frac{2\alpha}{m^2_c r^3}-\frac{\lambda}{2m^2_c r}\right)\vec{L}\cdot \vec{S}
+\frac{32\pi \alpha}{9m_c^2}\tilde{\delta}(r) \vec{S}_c\cdot \vec{S}_{\bar{c}}
+\frac{4\alpha}{m^2_c r^3}\left(\frac{\vec{S}_c\cdot \vec{S}_{\bar{c}}}{3}
+\frac{(\vec{S}_c\cdot\vec{r}) (\vec{S}_{\bar{c}}\cdot \vec{r})}{r^2}\right),
\label{finestructure}
\ee
where $\vec{L}$ denotes the relative orbital angular momentum,
$\vec{S}=\vec{S}_c+\vec{S}_{\bar{c}}$  is the total spin of the charm quark pairs
and $m_c$ is the charm quark mass.
The smeared delta function is taken to be
$\tilde{\delta}(r)=(\sigma/\sqrt{\pi})^3 e^{-\sigma^2 r^2}$~\cite{Barnes:2005pb,Li:2009ad}.
The Hamiltonian of the Schr\"{o}dinger equation in the quenched limit is represented as
\be\label{quenched}
H_0 =2m_c+\frac{p^2}{m_c}+V(r)+V_{s}(r).
\ee
We treat the spin-dependent term as a perturbation and
the spatial wave functions and bare mass $M_0$ are obtained by
solving the Schr\"{o}dinger  equation numerically using the Numerov method~\cite{Numerov:1927}.

\begin{table}[!htbp]
  \centering
\begin{tabular}{cccc}
\hline\hline
$\alpha$ & $\lambda$ & $c$ & $\sigma$\\
0.55 & $0.175~\text{GeV}^2$ &$-0.419~\text{GeV}$ & 1.45~GeV \\
\hline
$m_c$ & $m_s$ & $m_u$ & $m_d$\\
1.7 & 0.5 & 0.33 & 0.33\\
\hline\hline
$\psi(1S)$ & $\psi(2S)$ & $\psi(3S)$ & $\psi(4S)$\\
3.112 & 3.755 & 4.194 & 4.562 \\
\hline
$\psi(1D)$ & $\psi(2D)$ & $\psi(3D)$ & $\psi(4D)$\\
3.878 & 4.270 & 4.613 & 4.926\\
\hline\hline
\end{tabular}
\caption{Parameters of Cornell potential model and the corresponding bare mass spectrum.
The units of mass are GeV.}
\label{tab:4260Para}
\end{table}

The coupled channel effects show up when the explicit generation of the quark-antiquark pairs are considered.
We adopt the widely used \PPP/ model to generate the quark-antiquark pairs from the vacuum~\cite{LeYaouanc:1972ae,LeYaouanc:1973xz}.
In this model, the generated quark-antiquark pairs have the vacuum quantum numbers $J^{PC}=0^{++}$.
After simple arithmetic, one can conclude that the relative orbital angular momentum and the total spin are both equal to 1.
In the notation of ${}^{2S+1}L_J$, one should write it as \PPP/ which explains the name of this model.
For more information about the coupled channel effects and \PPP/ model,
see Ref.~\cite{Lu:2016mbb} and references therein.

The \PPP/ Hamiltonian can be expressed as
\be
H_I=2 m_q \gamma \int d^3x \bar{\psi}_q \psi_q,
\ee
where $m_q$ is the produced quark mass and $\gamma$ is the dimensionless coupling constant.
The $\psi_q$ ($\bar{\psi}_q$) is the spinor field to generate antiquark (quark).
Since the probability to generate heavier quarks should be suppressed,
we use the effective strength $\gamma_s=\frac{m_q}{m_s}\gamma$ in the following calculation,
where $m_q=m_u=m_d$ is the constituent quark mass of the up (or down) quark and $m_s$  is the strange quark mass.
The full Hamiltonian is $H=H_0+H_I$,
and the wave function of the physical state $\ket {A}$ is denoted as
\be
\ket {A}=c_0 \ket{\psi_0} +\sum_{BC} \int d^3p\, c_{BC}(p) \ket{BC;p},
\ee
where $c_0$ and $c_{BC}$ stand for the normalization constants of the bare state and the $BC$ components, respectively.
In this work, $B$ and $C$ refer to charmed and anticharmed mesons,
and the summation over $BC$ is carried out up to the ground state $P$ wave charmed mesons.
The effects from the $BC$ components are referred to as \cce/.
The mass shift caused by the $BC$ components and the probabilities of them are obtained after solving the Schr\"{o}dinger equation with the full Hamiltonian $H$.
They are expressed as
\bal
\Delta M&:=M-M_0=\sum_{BC}\int d^3p\, \frac{\vert \bra {BC;p} H_I \ket{\psi_0} \vert ^2}{M-E_{BC}-i\epsilon},\\
P_{BC}&:=\int d^3p|c_{BC}|^2=\int d^3p\, \frac{\vert \bra {BC;p} H_I \ket{\psi_0} \vert ^2}{(M-E_{BC})^2},\label{PBC}
\eal
where $M$ and $M_0$ are the eigenvalues of the full ($H$) and quenched Hamiltonian ($H_0$), respectively.
$E_{BC}=\sqrt{m_B^2+p^2}+\sqrt{m_C^2+p^2}$
and $P_{BC}$ is the \emph{unnormalized} probabilities, which is also called the coupling strength in next section.
In order to analyze different partial-wave contributions,
we adopt the Jacob-Wick formula to separate different partial waves of $P_{BC}$\cite{Jacob:1959at}.

The \cce/ calculation cannot proceed if the wave functions of the $\ket{\psi_0}$ and $BC$ components are not settled in Eq.(7).
Since the major part of the \cce/ calculation is encoded in the wave function overlap integration,
\be
\bra{BC;p} H_I \ket{\psi_0}=
\int d^3k
\phi_0(\vec{k}+\vec{p}) \phi_B^*(\vec{k}+x \vec{p})\phi_C^*(\vec{k}+x \vec{p})
|\vec{k}| Y_1^m(\theta_{\vec{k}},\phi_{\vec{k}}),\label{overlap}
\ee
where $x=m_q/(m_Q+m_q)$, and $m_Q$ and $m_q$ denote the charm quark and the light quark mass, respectively.
The $\phi_0, \phi_B$ and $\phi_C$ are the wave functions of $\ket{\psi_0}$ and $BC$ components, respectively and the notation $*$ stands for the complex conjugate.
These wave functions are in momentum space, and they are obtained by the Fourier transformation of the eigenfunctions of the bare Hamiltonian $H_0$.
Compared with the work in Ref.~\cite{Lu:2016mbb},
the \cce/ calculations of the \xy/ becomes more complicated due to the following reasons.

For the \dmeson/ components, the HQSS tells us that,
the experimentally observed $D_1(2420)$ (or $D_1$ in short) and $D_1(2430)$ (or $D_1'$) are not ${}^3P_1$ or ${}^1P_1$ states in the quark model, but are their linear combinations,
so the wave functions from the quark model should be modified accordingly.
The mixture can be formulated as
\be\label{HQSSMixing}
\left(
  \begin{array}{c}
   \ket{D_1}\\
   \ket{D_1'} \\
  \end{array}
  \right)=
\left(
  \begin{array}{cc}
    \cos\, \theta & \sin\, \theta\\
    -\sin\, \theta& \cos\, \theta\\
  \end{array}
  \right)
  \left(
  \begin{array}{c}
   \ket{{}^3P_1}\\
   \ket{{}^1P_1}\\
  \end{array}
  \right),
\ee
where $\theta$ is the mixing angle.
The HQSS predicts it to be $\theta_0=\arctan(\sqrt{2})\approx 54.7^\circ$,
which is called the ideal mixing angle \cite{Close:2005se,Qin:2016spb}.

Since for the heavy quarkonium, the heavy quarks in the initial states are treated as spectators,
the polarizations will not change after the generation of the quark-antiquark pairs,
we conclude that the \PPP/ model itself respects the HQSS.

Nevertheless, some HQSS breaking effects can still slip into the calculation of the \cce/.
The breaking effects lie in the input of the \dmeson/s,
which are reflected by the deviation from ideal mixing $\theta-\theta_0$
and the mass splitting of the charmed meson in a same $j_l$ multiplet,
where $j_l$ is the total spin of the light quarks in charmed mesons.
e.g., even though $D_2^*$ and $D_1$  belong to the same $j_l=3/2$ multiplet,
experimentally, they do not degenerate as claimed by the HQSS, revealing some HQSS breaking effect.

For \doned/ channel specifically,
we can conclude with or without respecting the HQSS by letting mixing angle $\theta$ freely run in the range $[0,\pi/2]$.
When $\theta$ reaches $\theta_0$, the HQSS is recovered.
For the other \dmeson/ channels, such as $D^*\bar{D}_2^*+c.c.$,
if the physical masses of the \dmeson/ are used,
the calculation will reflect some HQSS breaking effect,
and this case will be more realistic since the HQSS \emph{is} broken in reality.

For the charmonium component $\psi_0$ of the \xy/,
even though there is no room to accommodate the \xy/ in the quenched quark model,
\xy/ is in the mass range of $\psi(2D)$ to $\psi(3D)$ or $\psi(3S)$ to $\psi(4S)$\cite{Godfrey:1985xj,Segovia:2008zz}.
In order to compare the results under different assumptions on the bare states,
we calculate the \cce/ for all of the four states.
We borrow the parameters from Eq.~(2) in Ref.~\cite{Li:2009ad}
and the quenched masses predicted by $H_0$ in Eq.~(\ref{quenched}) are listed in Table~\ref{tab:4260Para}.

Note that, even though the mass predicted for $3D$ is nearly 350~MeV higher than \xy/ in Table~\ref{tab:4260Para},
the \cce/ can, in principle,
compensate such a large mass gap and shift the bare mass down to 4.26GeV\cite{Barnes:2007xu}.
Nevertheless, it is difficult to determine which bare state to shift and fitting the charmonium spectrum requires much {\it ab initio} calculation,
including the fixing of \PPP/'s coupling constant $\gamma$, which is beyond the scope of this paper.
So, in this work, $\gamma$ is not fixed,
and that is why $P_{BC}$ in Eq.~(\ref{PBC}) is unnormalized.
As a consequence, we cannot deduce the absolute probabilities of the charmonium core and the various \dmeson/ components in the \xy/,
however, by analyzing the $P_{BC}$, one can still draw some useful conclusions, as will be shown in the next section.

\section{Result Analysis}\label{sec:result}

\begin{figure}
  \centering
\includegraphics[width=\textwidth]{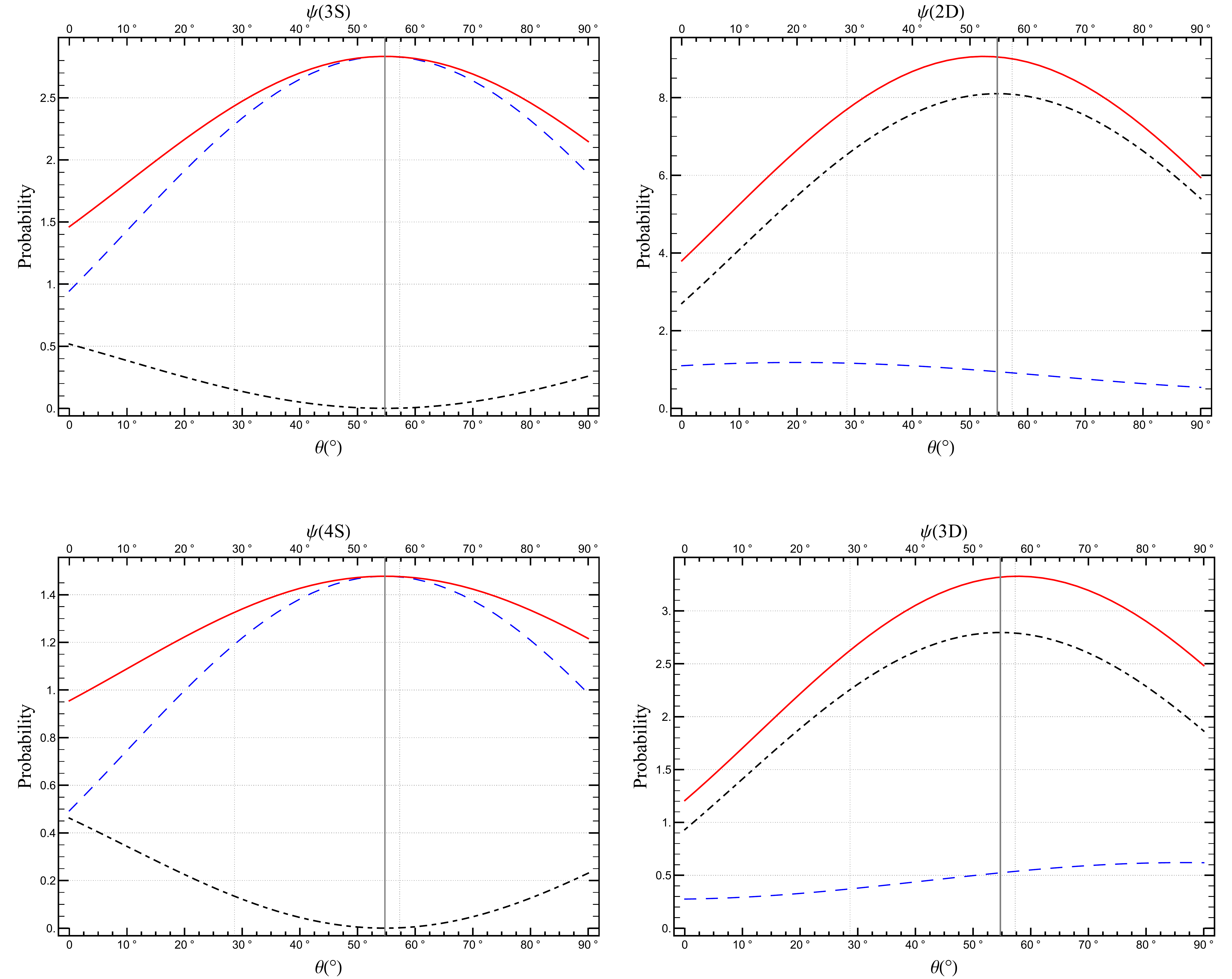}
  \caption{Partial wave probabilities of \doned/ for various initial states
  with respect to the mixing angle between $D_1({}^3P_1)$ and $D_1({}^1P_1)$ at 4251~MeV.
  The parameters are taken from Table~\ref{tab:4260Para}.
  \swave/, \dwave/ and total probability are depicted by black dot-dashed, blue dashed and red solid curves, respectively.
  The grey vertical line marks the ideal mixing $\theta_0\approx 54.7^\circ$.
  Only the relative amount has physical meaning because the \PPP/ model strength is not fixed.
}
  \label{fig:D1DRatio1}
\end{figure}

In order to explore the HQSS breaking effects,
we let the mixing angle run in the range $[0,\pi/2]$ and
plot the $S$ and $D$ wave couplings to \doned/ for the different charmonium states in \autoref{fig:D1DRatio1}.

Figure~\ref{fig:D1DRatio1} clearly shows that \doned/ cannot couple to the $\psi(3S)$ or $\psi(4S)$ in the \swave/ in the HQSS limit,
and this ideal mixing will enhance their $D$ wave couplings to \doned/.
For the $\psi(2D)$ and $\psi(3D)$, both the $S$ and $D$ wave couplings are allowed,
with the \swave/ probability around 7 times larger than the \dwave/ one at $\theta_0$.

The sum of both $S$ and $D$ wave couplings are shown in Table~\ref{tab:D1Compare},
where the results using the parameters in Ref.~\cite{Barnes:2007xu} are also listed for comparison.
One can clearly see that the \doned/ couples stronger to $^3D_1$ than to the $^3S_1$ charmonia.
This conclusion is somewhat model independent since it comes from two different sets of parameters.

The \doned/ molecule scenario assumes the long-distance component of \xy/ to be \doned/,
the short-distance $c\bar{c}$ core is not specified.
Our calculation suggests that this short distance charmonium core should be dominantly $\psi(nD)$ rather than $\psi(nS)$.

If the charmonium core of \xy/ is $\psi(nD)$, the wave function at the origin is zero.
Since the $\psi(nD)$ can only couple to the virtual photon at the next-to-next-to leading order \cite{Rosner:2001nm},
its direct production is suppressed at the $e^+e^-$ collider.
The $R$ ratio measured by the BES Collaboration\cite{Ablikim:2007gd} reveals that
the total cross section has a dip instead of a peak around $4.26~\text{GeV}$.
The picture that the \xy/ has a $\psi(nD)$ core is in agreement with this experimental fact.
In this sense, our calculation supports the \doned/ scenario.

\begin{table}[!hbtp]
  \centering
\begin{tabular}{c|c|ccccccccccccc}
\hline\hline
& Coupled channels & $\psi(3S)$ & $\psi(4S)$  & $\psi(2D)$  & $\psi(3D)$\\
\hline
\multirow{2}*{Parameters in Table~\ref{tab:4260Para}}
& \doned/ & 2.83 & 1.48  & 9.05  & 3.32 \\
& $D_1'\bar{D}$ & 0.75 & 0.62  & 0.68  & 0.36 \\
\hline
\multirow{2}*{Parameters in Ref.\cite{Barnes:2007xu}}
& \doned/ & 1.05 & 0.33  & 5.17  & 1.09 \\
& $D_1'\bar{D}$ & 1.44 & 0.54  & 0.69  & 0.37 \\
\hline\hline
\end{tabular}
\caption{Coupling strength of the \doned/ and $D_1'\bar{D}$ channels for the different charmonium states in the HQSS limit.}
\label{tab:D1Compare}
\end{table}

\begin{table} [!htbp]
  \centering
\begin{tabular}{l|llll|lllllll}
\hline\hline
\multirow{2}*{Coupled Channels}&\multicolumn{4}{c|}{Parameters in \autoref{tab:4260Para}} &\multicolumn{4}{c}{Parameters in Ref.\cite{Barnes:2007xu}} \\
 \cline{2-9}
& $\psi(3S)$ & $\psi(4S)$ & $\psi(2D)$ & $\psi(3D)$ & $\psi(3S)$ & $\psi(4S)$ & $\psi(2D)$ & $\psi(3D)$\\
\hline
$D-D^*_0$ &   0& 0& 0& 0& 0& 0& 0& 0\\
 $D^*-D^*_0$ &  0.215 & 0.238 & 0.028 & 0.07 & 0.809 & 1.073 & 0.073 & 0.167 \\
 $D-D_1'$ &  0.265 & 0.42 & 0.076 & 0.109 & 1.367 & 1.635 & 0.133 & 0.336 \\
 $D^*-D_1'$ &  0.325 & 0.317 & 0.075 & 0.118 & 1.117 & 1.651 & 0.148 & 0.378 \\
 $\mathbf{D-D_1}$ &  1 & 1 & 1 & 1 & 1 & 1 & 1 & 1 \\
 $D^*-D_1$ &  0.616 & 0.564 & 0.149 & 0.18 & 0.816 & 0.922 & 0.185 & 0.269 \\
 $D-D^*_2$ &  0.629 & 0.59 & 0.065 & 0.093 & 0.7 & 0.716 & 0.069 & 0.1 \\
 $D^*-D^*_2$ &  0.992 & 0.914 & 0.225 & 0.339 & 1.384 & 1.632 & 0.267 & 0.537 \\
\hline
 $D_s-D^*_{s0}$ &  0& 0& 0& 0& 0& 0& 0& 0\\
 $D_s^*-D^*_{s0}$ &  0.043 & 0.041 & 0.004 & 0.01 & 0.11 & 0.168 & 0.01 & 0.024 \\
 $D_s-D_{s1}(2536)$ &  0.035 & 0.035 & 0.013 & 0.02 & 0.095 & 0.151 & 0.023 & 0.06 \\
 $D_s^*-D_{s1}(2536)$ &  0.054 & 0.056 & 0.014 & 0.022 & 0.16 & 0.27 & 0.024 & 0.066 \\
 $D_s-D_{s1}(2460)$ &  0.08 & 0.052 & 0.033 & 0.027 & 0.1 & 0.118 & 0.039 & 0.054 \\
 $D_s^*-D_{s1}(2460)$ &  0.089 & 0.066 & 0.02 & 0.021 & 0.132 & 0.178 & 0.028 & 0.05 \\
 $D_s-D^*_{s2}$ &  0.05 & 0.036 & 0.005 & 0.005 & 0.071 & 0.094 & 0.006 & 0.011 \\
 $D_s^*-D^*_{s2}$ &  0.129 & 0.102 & 0.037 & 0.047 & 0.209 & 0.306 & 0.052 & 0.119 \\
\hline
 $D_1'-D_1'$ &  0.155 & 0.122 & 0.059 & 0.07 & 0.273 & 0.513 & 0.102 & 0.226 \\
 $D_1'-D^*_0$ &  0.072 & 0.045 & 0.006 & 0.005 & 0.097 & 0.175 & 0.011 & 0.024 \\
 $D_1'-D_1$ & 0.128 & 0.112 & 0.023 & 0.035 & 0.225 & 0.434 & 0.034 & 0.105 \\
 $D_1'-D^*_2$ &  0.206 & 0.266 & 0.044 & 0.083 & 0.566 & 1.077 & 0.094 & 0.277 \\
 $D^*_0-D^*_0$ &  0.027 & 0.018 & 0.016 & 0.017 & 0.034 & 0.062 & 0.027 & 0.064 \\
 $D^*_0-D_1$ & 0.052 & 0.042 & 0.01 & 0.012 & 0.08 & 0.159 & 0.017 & 0.043 \\
 $D^*_0-D^*_2$ &  0.172 & 0.12 & 0.045 & 0.055 & 0.273 & 0.522 & 0.046 & 0.19 \\
 $D_1-D_1$ &  0.16 & 0.154 & 0.049 & 0.069 & 0.331 & 0.614 & 0.097 & 0.211 \\
 $D_1-D^*_2$ &  0.264 & 0.21 & 0.067 & 0.081 & 0.48 & 0.922 & 0.1 & 0.265 \\
 $D^*_2-D^*_2$ &  0.157 & 0.23 & 0.053 & 0.119 & 0.5 & 0.918 & 0.104 & 0.324 \\
\hline
 $D_{s1}(2536)-D_{s1}(2536)$ &  0.027 & 0.026 & 0.011 & 0.014 & 0.062 & 0.124 & 0.019 & 0.045 \\
 $D_{s1}(2536)-D^*_{s0}$ &  0.013 & 0.009 & 0.001 & 0.001 & 0.02 & 0.037 & 0.001 & 0.003 \\
 $D_{s1}(2536)-D_{s1}(2460)$ &  0.028 & 0.027 & 0.005 & 0.009 & 0.059 & 0.122 & 0.009 & 0.03 \\
 $D_{s1}(2536)-D^*_{s2}$ &  0.051 & 0.063 & 0.011 & 0.021 & 0.147 & 0.306 & 0.023 & 0.075 \\
 $D^*_{s0}-D^*_{s0}$ &  0.007 & 0.005 & 0.004 & 0.005 & 0.01 & 0.018 & 0.007 & 0.017 \\
 $D^*_{s0}-D_{s1}(2460)$ &  0.014 & 0.013 & 0.003 & 0.004 & 0.027 & 0.056 & 0.005 & 0.013 \\
 $D^*_{s0}-D^*_{s2}$ &  0.034 & 0.029 & 0.01 & 0.015 & 0.07 & 0.14 & 0.016 & 0.051 \\
 $D_{s1}(2460)-D_{s1}(2460)$ &  0.038 & 0.038 & 0.01 & 0.014 & 0.087 & 0.168 & 0.018 & 0.041 \\
 $D_{s1}(2460)-D^*_{s2}$ &  0.053 & 0.05 & 0.014 & 0.02 & 0.121 & 0.246 & 0.024 & 0.067 \\
 $D^*_{s2}-D^*_{s2}$ &  0.042 & 0.054 & 0.015 & 0.028 & 0.12 & 0.24 & 0.029 & 0.086 \\
\hline\hline
\end{tabular}
\caption{Coupling strength of various coupled channels divided by \doned/ ratio in the ideal mixing case,
where $\bar{D}$ meson is represented by $D$ meson for clarity and the charge conjugation is always implied.
 i.e., the $D-D_1$ stands for the channel $D\bar{D}_1+c.c.$.
One should not compare the numbers between different columns until \doned/ values in Table~\ref{tab:D1Compare} are multiplied.
}
\label{tab:4260CoupleRatio}
\end{table}

It is also reasonable to ask whether this conclusion will change if HQSS is broken,
since charm quark is not infinitely heavy.
Experimentally, the deviation from ideal mixing is ($5.7^\circ\pm 4^\circ$)\cite{Abe:2003zm},
which is very small compared with $\theta_0\approx 54.7^\circ$.
When this deviation is added to the $\theta_0$, the mixing angle becomes $\theta\approx 60.4^\circ$.
Since the red curve in \autoref{fig:D1DRatio1} reaches the maximum around $60.4^\circ$,
the previous conclusion is still correct.

For the \doned/ scenario, there is also one more benefit to choose a $\psi(nD)$ core
--- the coupling to \doned/ will be larger than any other \dmeson/ components,
which will not always be the case for a $\psi(nS)$ core.
To see this, we list the probabilities of all the \dmeson/s divided by that of \doned/ in Table~\ref{tab:4260CoupleRatio}.

As is explained in the previous section,
the breaking effect of the HQSS is reflected by the mass splitting of same $j_l$ multiplets, such as $D_2^*$ and $D_1$,
so the results in Table~\ref{tab:4260CoupleRatio} do not respect the HQSS.
However, the results are more valuable since, in the various \dmeson/ molecule scenarios,
different physical masses are applied to the \dmeson/s.
In order to become fully realistic,
the mixing angle $\theta$ in Table~\ref{tab:4260CoupleRatio} should be around $60.4^\circ$.
However, since the coupling strength to \doned/ barely changes around $\theta_0$,
the modification to Table~\ref{tab:4260CoupleRatio} is negligible.

The couplings to the channels that are far above 4.26 GeV are generally small.
This is a universal conclusion from the \cce/.
From Eq.~(7), one can readily see that the asymptotic behavior of $P_{BC}$ is proportional to $1/(m_B+m_C)^2$.
If the coupled channels are father from the \xy/, their contributions will be naturally suppressed (mass suppression mechanism for short).
For the $D_s$ meson channel, an additional suppression comes from the effective strength of \PPP/ model $\gamma_s$.
Since $\gamma_s\approx0.66 \gamma$, the couplings to $D_s$ mesons are universally smaller than the $c\bar{u}$ or $c\bar{d}$.

For the coupled channels where meson pairs are $D(1S)$ and $\bar{D}(1P)$,
if we respect the HQSS and set $m(D)=m(D^*)$ and $m(D_0^*)=m(D_1)=m(D_1')=m(D_2^*)$,
the contribution of all the eight channels (corresponding to the first eight rows in Table~\ref{tab:4260CoupleRatio}) will be the same order of magnitude (except for the $DD_0^*$ channel, it is forbidden by the conservation of angular momentum and parity).
In this case, the largest coupling will come from the $D^*D_2^*$ channel,
since the spin configurations of this channel are more than the others (spin enhancement mechanism for short).
However, if the physical masses are applied to these \dmeson/s,
the $D^*D_2^*$ channel will be father from the \xy/,
and contributions from this channel will be suppressed.
In a short summary, for the realistic case,
the two mechanisms, mass suppression and spin enhancement,
have to compete with each other to tell which channel gives the dominant contribution.

Our calculation shows that if the charmonium core is $\psi(nD)$,
the mass suppression mechanism will overtake.
Since the \doned/ channel are closer to \xy/,
the contributions from the non-\doned/ components will be highly suppressed,
which makes the name of the ``\doned/" molecule more reasonable.
In contrast, if the charmonium core is $\psi(nS)$,
the two mechanisms will be just as importance with each other,
as a consequence, other molecules will have non-negligible contributions.
For example, in Table~\ref{tab:4260CoupleRatio}, the coupling to the $D^*\bar{D}^*_2+c.c.$  is very close to the \doned/,
and in the parameters of Ref.~\cite{Barnes:2007xu},
this coupling even exceeds the \doned/'s,
although their simple harmonic oscillator wave functions are more inaccurate.
When this happens, it will be better to call the \xy/ a $D^*\bar{D}^*_2$ rather than a \doned/ molecule.

Finally, we need to stress that, even though the non-\doned/ components are suppressed when the charmonium core is $\psi(nD)$, the contribution of these components could still be sizable.
As shown in Table~\ref{tab:4260CoupleRatio},
for $\psi(2D)$, the contribution from the $D^*D_2^*$ channel is still around $1/4$ of the \doned/.
The impacts of these extra \dmeson/ components are still worth studying.

\section{Summary}\label{sec:summary}

We calculated the probabilities of various \dmeson/ molecular components for $\psi(3S)$, $\psi(4S)$, $\psi(2D)$ and $\psi(3D)$
under the \PPP/ framework.
Our calculation reveals that, even though heavy quark spin symmetry forbids \swave/ coupling of \doned/ to ${}^3S_1$ charmonia,
the \dwave/ coupling is allowed and not small.

The more interesting result is that the \doned/ couples more strongly to the ${}^3D_1$ charmonia.
This means that the short distance charmonium core of \xy/ should be dominantly $\psi(nD)$ other than $\psi(nS)$ in the \doned/ molecular scenario.
This $\psi(nD)$ core of the \xy/ agrees with the experimental fact that the $R$ ratio has a dip around 4.26~GeV.
In this sense, our calculation supports the \doned/ molecular scenario of the \xy/.

Choosing a $\psi(nD)$ core will also suppress the probabilities of the non-\doned/ meson molecules,
making the \doned/ molecule picture more distinctive.
Nevertheless, even though these non-\doned/ components are suppressed,
their contributions may be not negligible.
The impacts of these extra \dmeson/ components still remain to be explored.

\section*{Acknowledgements}
We are grateful to Feng-Kun Guo, Qian~Wang, and Qiang~Zhao for various discussions and suggestions.
This work is supported by the National Natural Science Foundation of China under Grants No.~11621131001 (CRC110 by DFG and NSFC) and No.~11647601.
M. Naeem~Anwar is supported by CAS-TWAS President's Fellowship for International Ph.D. Students.

\providecommand{\href}[2]{#2}\begingroup\raggedright\endgroup

\end{document}